\documentclass[3p,times]{elsarticle}

\usepackage{ecrc}
\usepackage{wrapfig}

\usepackage{epstopdf}

\volume{00}

\firstpage{1}

\journalname{Nuclear Physics A}

\runauth{Taku Gunji et al.}

\jid{nupha}

\jnltitlelogo{Nuclear Physics A}

\usepackage{graphicx}
\usepackage{amsmath,amssymb}

\begin{document}

\begin{frontmatter}



\title{Future Upgrade and Physics Perspectives of the ALICE TPC}

\author{Taku Gunji (for the ALICE\fnref{col1} Collaboration)}
\fntext[col1] {A list of members of the ALICE Collaboration and acknowledgements can be found at the end of this issue.}
\address{Center for Nuclear Study, Graduate School of Science, University of Tokyo, 7-3-1 Hongo, Bunkyo, Tokyo 113-0033, Japan}

\begin{abstract}
The ALICE experiment at the Large Hadron Collider (LHC) proposes 
major detector upgrades to fully exploit the increase of the luminosity of the LHC in RUN~3 
and to extend the physics reach for rare probes at low transverse momentum.
The Time Projection Chamber (TPC) is one of the main tracking and PID devices in the central 
barrel of ALICE. The maximum trigger rate of the TPC is currently limited to about 3.5 kHz 
by the operation of a gating grid system.
In order to make full use of the luminosity in RUN 3, the TPC 
is foreseen to be operated in an ungated mode with continuous readout. 
The existing MWPC readout will be replaced by a 
Micro-Pattern Gaseous Detector (MPGD) based readout, 
which provides intrinsic ion capture capability without gating.
Extensive detector R\&D employing Gas Electron Multiplier (GEM) 
and Micro-Mesh Gaseous detector (Micromegas) technologies, and simulation 
studies to advance the techniques for 
the corrections of space-charge distortions have been performed 
since 2012. In this paper, the expected detector performance 
and the status of the R\&D program to achieve this ambitious goal are described.

\end{abstract}


\end{frontmatter}


\section{ALICE Upgrade after LS2}
\label{alice_upgrade}
The ALICE experiment is dedicated to the studies of the properties of the 
deconfined QCD medium (Quark-Gluon Plasma, QGP) by conducting  
ultra-relativistic heavy-ion collisions at the LHC~\cite{bib:ref1}.
A significant increase of the luminosity for heavy ions is expected in RUN 3 after Long 
Shutdown 2 (LS2), which implies a collision rate of about 50 kHz 
and $\mathcal{L}_{\rm int}$ = 10 nb$^{-1}$.
This luminosity upgrade provides a substantial enhancement of capabilities for
measuring observables relevant to the characterization of the QGP at the
highest temperatures~\cite{bib:ref2_loi}.

In order to exploit the scientific potential of the high-luminosity heavy-ion program in RUN~3, 
ALICE plans to extend its physics reach by upgrading the ALICE detector. 
The major goals of the upgraded ALICE detector are as follows; 
precision measurements of heavy-quark and quarkonia production at low transverse 
momentum ($p_{\rm T}$) to study the mechanisms of heavy-quark thermalization 
and interactions in the medium,  production of low-mass dielectrons to extract 
information on thermodynamical properties of the medium and to characterize 
the chiral phase transition, jets and jet correlations to reveal the mechanisms 
of partonic energy loss in the medium~\cite{bib:ref2_loi}.


\section{ALICE TPC Upgrade}
\label{tpc_ugrade}
The Time Projection Chamber (TPC) is one of the main tracking and PID devices in the central 
barrel of the ALICE detector. It provides precise charged-particle tracking, 
momentum measurement, and particle identification in very high multiplicity 
heavy-ion collisions~\cite{bib:ref3}. 

The readout rate of the TPC is currently limited by the necessity to prevent ions
generated in the the amplification region of the MWPC-based readout chambers
from drifting back into the drift volume, 
which is achieved through active ion gating by operating a dedicated gating grid. 
The relevant ion drift times limit the maximum trigger rate of the TPC to about 3.5 kHz.

Operation of the current TPC with the MPWC-based readout scheme and the current 
active ion gating scheme at 50 kHz Pb-Pb collisions in RUN~3 cannot be possible.
On the other hand, operation of the current TPC with continuously open gating grid
cannot be the solution since back-drifting ions from the amplification region will lead to 
excessive ion charge densities and distortions of the electric field in the drift volume.
The proposed scheme to acquire high rate operational capability and a small number of  
back-drifting ions is to replace the existing MWPC-based readout chambers and gating grid system 
by a multi-layer Gas Electron Multiplier (GEM) system and to run the TPC in an 
ungated continuous mode.  GEMs have been developed to 
cope with the stringent requirements for high-luminosity experiments~\cite{bib:ref4}
and have proven to provide excellent position resolution, to have very high rate capability, 
and better ion blocking capability compared to MWPC.
The main considerations for the TPC upgrade and the design requirements are as follows~\cite{bib:ref5}:
 \begin{itemize}
 \item{The maximum ion backflow (IBF) that can be tolerated is about 1\% at a gain of 2000 in 
 	Ne-CO$_2$-N$_2$ (90-10-5), i.e. 20 back-drifting ions per incoming primary electron ($\epsilon$ = 20).}
\item{In case of IBF = 1\%,  space-charge field distortions reach 20~cm 
  and 8~cm in $r$ and $r\phi$ at small $r$ and $z$ ($|\eta|\sim0$) in the TPC, respectively. 
  In order to preserve the present momentum resolution, online and offline distortion 
  corrections with a precision better than 500~$\mu$m, i.e. a few times 10$^{-3}$, 
  are required.}
\item{Due to the limited bandwidth of the data 	acquisition system, 
  reduction of data flow size by a factor of 20 is needed in the online 
  reconstruction 
  by finding the clusters associated to tracks.}
\item{The upgraded TPC must preserve the performance of the existing system in terms of 
  particle identification via d$E$/d$x$, implying a local energy resolution better than 
  12\% (at 5.9 keV).}
\end{itemize}

\section{Status of R\&D Activities}
\label{rd_status}
An extensive R\&D program has been started in 2012 to study the performance of GEM-based detectors 
(IBF, gain stability, discharge probability), technology choice (GEM stacks including 
the combination of GEMs with different pitches,  COBRA-GEM,  2 GEM + Micromegas system), 
large prototype production by single mask technology, electronics R\&D, 
and simulation studies to establish the strategy for space-charge distortion corrections.

\begin{figure}[htbp]        
\begin{minipage}{0.5\hsize}             
\begin{center}                                                                                                                                                  
\includegraphics*[width=8.75cm]{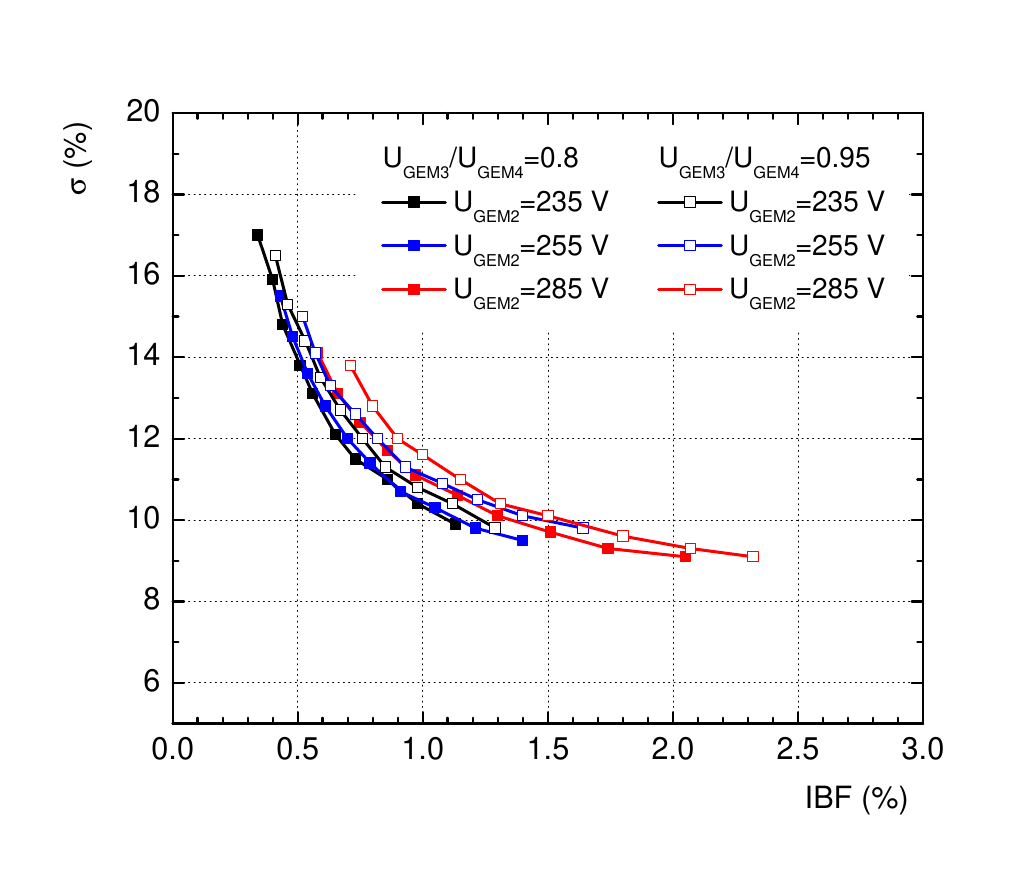}                                                                                                     
\caption{                                                                                                                                                       
       Correlation between IBF and energy resolution at 5.9 keV in a                                                                                   
       4 GEM setup (S-LP-LP-S)  in Ne-CO$_2$-N$_2$ (90-10-5) for various settings of                                                                            
       voltage of GEM2.}                                                                                                                                        
\label{fig:fig3}                                                                                                                                                
\end{center}                                                                                                                                                    
\end{minipage}                                                                                                                                                  
\begin{minipage}{0.5\hsize}
\begin{center}
\includegraphics*[width=6.7cm]{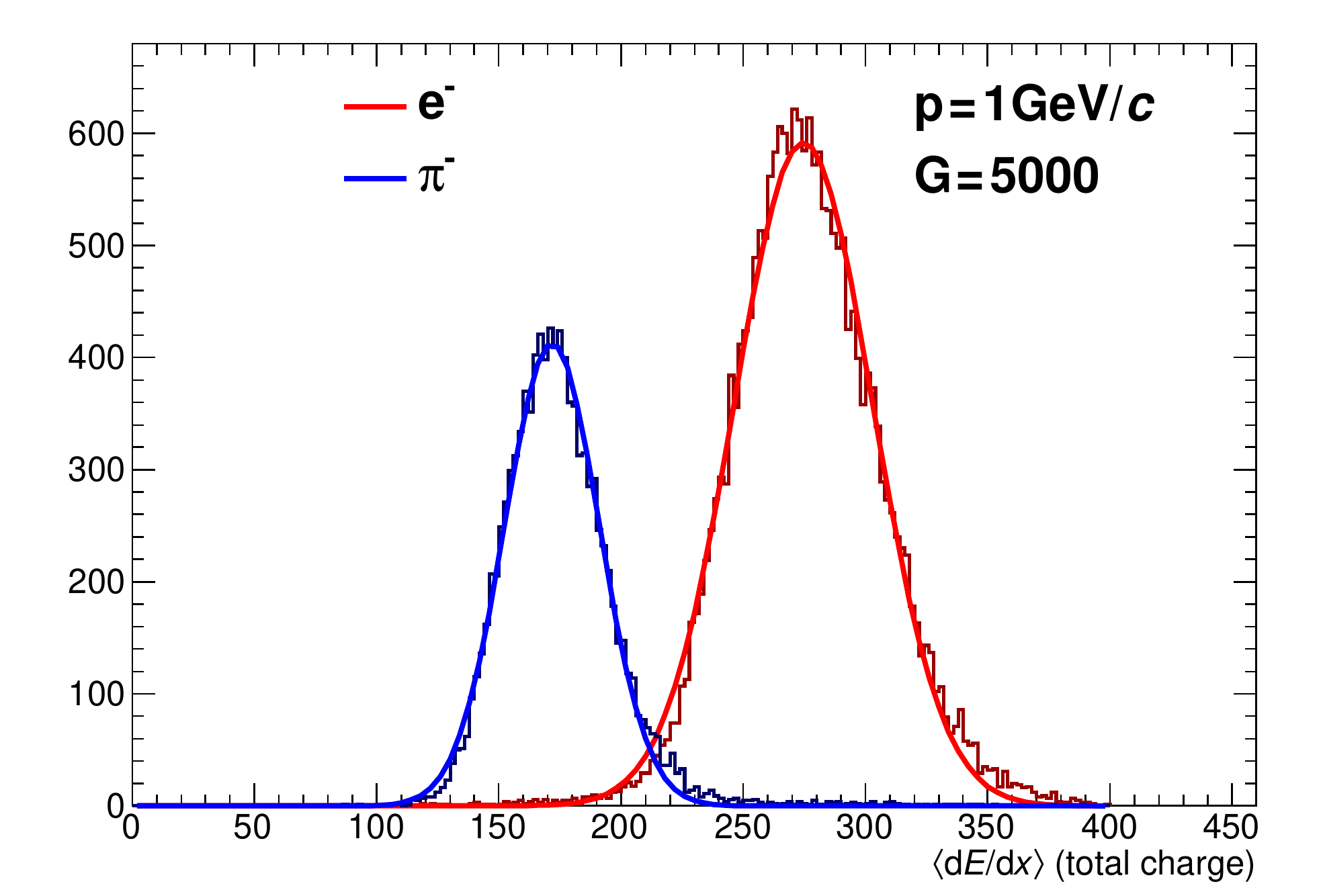}
\caption{
	d$E$/d$x$ spectrum of 1~GeV/$c$ electrons and pions recorded at a gain of $\sim$ 5000 
	measured with an IROC prototype employing a triple stack of large-size GEM foils. (Y-axis
	shows the number of counts and X-axis shows d$E$/d$x$ (a.u.).}
\label{fig:fig4}
\end{center}
\end{minipage}
\end{figure}

Our baseline solution comprises stacks of 4 GEM layers, where 1st and 4th GEMs are standard GEMs with 140~$\mu$m pitch, 
50~$\mu$m thickness, and 70 (50)~$\mu$m outer (inner) hole diameter, and 2nd and 3rd GEMs are large pitch GEM foils with 
280~$\mu$m pitch, 50~$\mu$m thickness, and 70 (50)~$\mu$m outer (inner) hole diameter.
This setup allows to block ions efficiently by 
employing 
low/high fields above/below GEMs and foils with low optical transparency.
Figure~\ref{fig:fig3} shows the results of the measured IBF and 
energy resolution at 5.9 keV at a gain of 2000 for a 4-GEM system, where 
the voltage across GEM1 increases from left to right along the x-axis.
It can be seen that an IBF of 0.7\% is achieved at an energy resolution of 12\% (at 5.9 keV)
The observed anti-correlation between IBF and resolution is related to the gains of the first 
two GEMs: higher electron multiplication at the early stages improves the energy resolution, 
while it results in larger number of ions escaping into the drift region. 

Detailed simulations based on Garfield++~\cite{bib:ref6} were performed to describe the observed 
IBF performance. It was found that IBF is very sensitive to the alignment of the GEM holes 
in consecutive layers, which can not be controlled experimentally. 
The measured IBF values are best reproduced in simulations, if a random misalignment of the 
holes is assumed, corresponding to the most probable relative geometrical position of 
GEM foils in a stack.

A prototype of an Inner Readout Chamber of the TPC (IROC) was built in 2012, where triple stacks 
of GEMs were produced using the single-mask technology developed by the MPGD workshop at CERN. 
Beam test was carried out  
at the PS-T10 beamline and the d$E$/d$x$ resolution was studied as function of the transfer fields 
and voltages across GEMs. Figure~\ref{fig:fig4} shows the d$E$/d$x$ spectra of 1~GeV/$c$ 
electrons and pions recorded at a gain of $\sim$ 5000. The energy resolution is 10.5\% for the 
IBF-optimized field configurations and the resolution is comparable to the d$E$/d$x$ resolution 
of the current TPC.

\begin{wrapfigure}{r}{73mm}
  \begin{center}
    \includegraphics*[width=7.3cm]{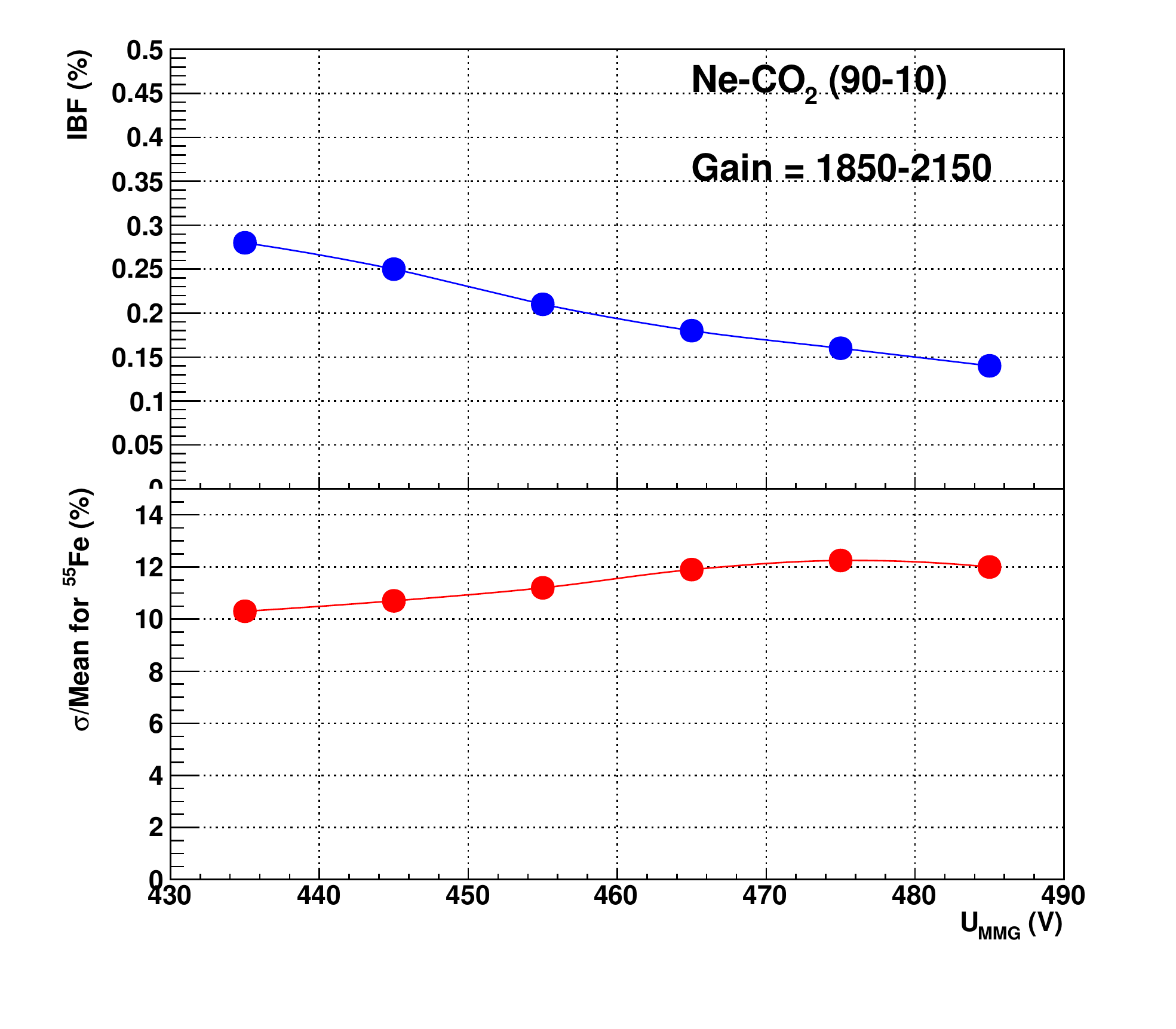}
    \caption{
	Upper: IBF as a function of voltage at mesh for 2 GEM + MM system. 
	Bottom: Energy resolution as a function of voltage at mesh for 2 GEM + MM system 
	}
\label{fig:fig5}
\end{center}
\end{wrapfigure}

An alternative solution is 
a system combining 2 GEMs 
with a Micromegas detector. Micromegas (MM) provides low IBF 
due to the larger ratio of the electric field values in the small amplification gap 
to the drift field above the MM. If the MM employs a fine mesh (400-1000 LPI),  IBF is 
close to the ratio between two fields itself~\cite{bib:ref7}.
The IBF and energy resolution for this hybrid 2-GEM + MM system were measured, using a 10 $\times$ 10 cm$^2$ prototype detector. 
The results are shown in Fig.~\ref{fig:fig5}, 
where an IBF of ~0.2\% is reached at an energy resolution of 12\% at 5.9 keV.
A large-scale solution for the inner and outer TPC readout chambers 
and the operational stability will be verified in the future. 

The new scheme also required the development of new front-end electronics 
to cope with the reversed 
polarity, the requirements for continuous readout, and the increased data throughput in high rate Pb-Pb collisions.
A new front-end ASIC called SAMPA has been developed, which 
integrates the functionality of the present preamp/shaper and ALTRO ADC+DSP (Digital Signal Processing)
and supports continuous or triggered readout~\cite{bib:ref6, bib:ref8}.
First MPW (Multi-Project Wafer) submission was done in April 2013 and further developments 
are ongoing.

\begin{figure}[htbp]
\begin{center}
\includegraphics*[width=15.cm]{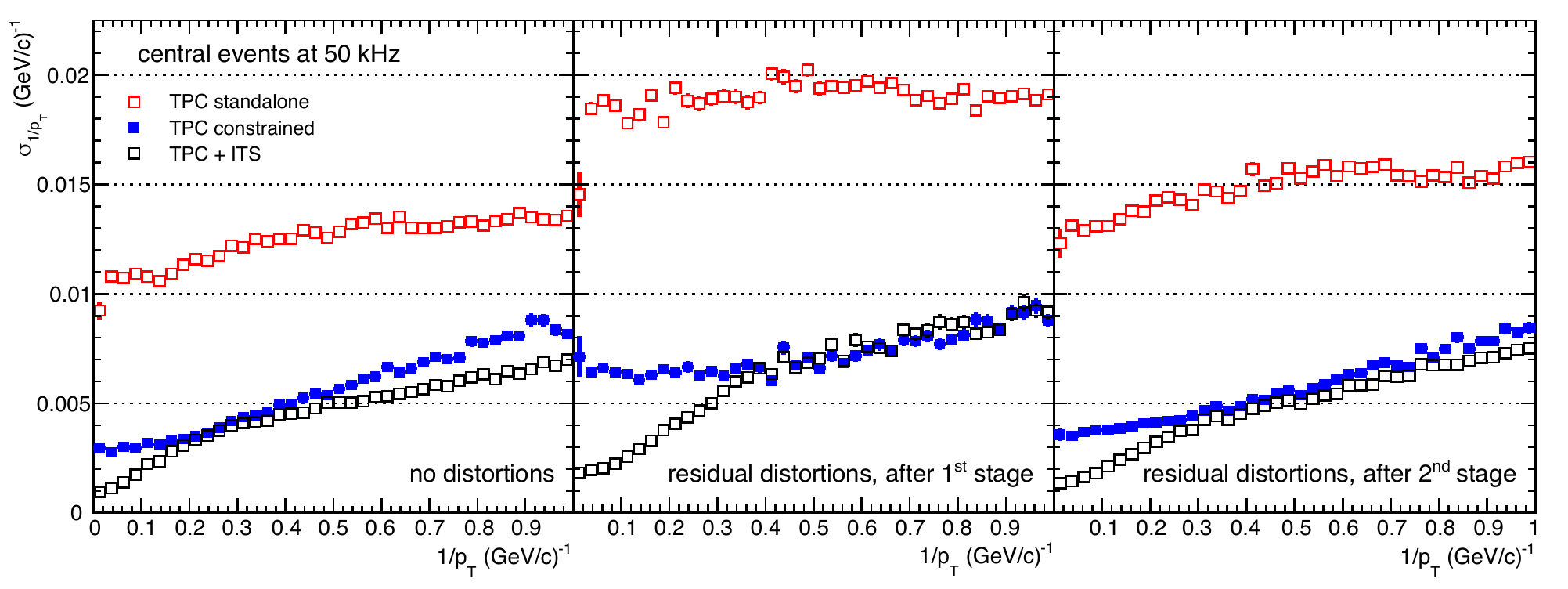}
\caption{
   Comparison of the momentum resolution without distortions
   (left) and with residual distortions after the first
   (middle) and second (right) reconstruction stage in Pb-Pb
   collisions at 50 kHz}
\label{fig:fig6}
\end{center}
\end{figure}

Online and offline reconstruction and calibration are very challenging 
due to the demand of data compression and 
requirement of the space-charge distortion corrections. 
Currently a two-stage reconstruction scheme is under consideration.  
In the first stage of the reconstruction, an averaged space-charge distortion map 
scaled to the averaged multiplicity for certain time intervals is used for the distortion corrections, 
and cluster finding and cluster-track association are performed, which leads to a data compression by a factor of 20.
Full tracking with the external detectors (Inner Tracking System + Transition Radiation Detector) will be performed in the 2nd stage of the reconstruction, 
where a high-resolution space-charge map being updated every 5~msec is generated for the full 
distortion corrections.  Figure~\ref{fig:fig6} shows the expected $p_{\rm T}$ resolution 
in 50 kHz Pb-Pb collisions without 
any space-charge distortions (left), with space-charge distortion and 
distortion corrections at the first stage (middle), and with space-charge distortion and 
full distortion corrections at 2nd stage of reconstruction (right). In these calculations, 
space-charge fluctuations mainly due to the number of pileup events and 
charged particle multiplicities are taken into account. 
The obtained $p_{\rm T}$ resolution after the 2nd reconstruction stage 
is comparable to that without distortions, 
if TPC-ITS global tracks are considered.

\section{Summary and Outlook}
\label{summary}
To exploit the full potential of the high luminosity of the LHC in RUN 3, the ALICE program 
for RUN 3 requires an upgrade of the TPC.
The heart of the TPC upgrade is to replace the MWPC-based readout chambers by 
detectors employing micro-pattern detectors including GEMs to allow TPC operation in continuous mode.
Extensive detector R\&D and simulations have been conducted and a baseline scenario 
for the detector design has been established. 
Quadruple stacks of GEM layers with different GEM pitches provide the required IBF and energy resolution. 
Also a design based on a hybrid configuration of GEMs and Micromegas is studied.
Simulations show that the performance of the present TPC can be retained in 50 kHz 
Pb-Pb collisions after distortion corrections. 
Further studies of the long-term stability, uniformity of the gain and IBF, 
and discharge probability are being conducted. 
IROC prototypes employing a 4-GEM stack and a hybrid 2-GEM + MM system are being built and 
beam tests will be carried out at the PS and the SPS in the fall of 2014.








\end{document}